\documentclass[aps,pra,twocolumn,amssymb]{revtex4}
\input epsf.tex

\begin{document}

\title{A Relevant Two Qubit Bell Inequality Inequivalent to the CHSH Inequality}
\author{Daniel Collins and Nicolas Gisin}
\affiliation{Group of Applied Physics, University of Geneva, 20,
rue de l'Ecole-de-M\'{e}decine, CH-1211 Geneva 4, Switzerland}

\date{28 August 2003}

\begin{abstract}

We computationally investigate the complete polytope of Bell
inequalities for 2 particles with small numbers of possible
measurements and outcomes.  Our approach is limited by Pitowsky's
connection of this problem to the computationally hard NP problem.
Despite this, we find that there are very few relevant
inequivalent inequalities for small numbers.  For example, in the
case with 3 possible 2-outcome measurements on each particle,
there is just one new inequality.  We describe mixed 2-qubit
states which violate this inequality but not the CHSH.  The new
inequality also illustrates a sharing of bi-partite non-locality
between three qubits: something not seen using the CHSH
inequality.  It also inspires us to discover a class of Bell
inequalities with m possible n-outcome measurements on each
particle.

\end{abstract}

\maketitle

\newcommand{\ket}[1]{\left | #1 \right \rangle}
\newcommand{\bra}[1]{\left \langle #1 \right |}
\newcommand{\amp}[2]{\left \langle #1 \left | #2 \right. \right \rangle}
\newcommand{\proj}[1]{\ket{#1} \! \bra{#1}}


How are actions and events in different places connected to one
another? Normally we imagine that the correlations were arranged
in the past.  Both my socks are black since I put on a pair this
morning. However our quantum mechanical theory of the world is
more complicated \cite{Bell}. Correlations are created in at least
one more way. Some possibilities are that a) correlations are
arranged through faster than light influences in the present, b)
correlations are arranged in a many world scenario, c)
correlations just occur: they constitute a primary concept,
preventing us from consistently thinking about local subsystems.
Since these possibilities are disliked by many physicists, we have
studied the set of correlations which can be generated by the
past, and when quantum mechanics goes beyond this.  Our goal is
first to find a simple set of conditions - generalised Bell
inequalities - which describe the boundaries of the set of
past-generated correlations. This set is often called the set of
common cause correlations, local hidden variable (lhv)
correlations, local variable correlations, or local realistic
correlations.  Our second goal is to see when quantum mechanics
goes outside this set.

One reason for studying this boundary so closely is the
fundamental question: which correlations can be generated in this
way?  A second reason is that a violation of Bell inequalities
gives a signature for useful entanglement. For instance, violation
of a certain Bell inequality by an N qubit state implies that the
state is distillable \cite{belldistill}: perfect bipartite
entanglement can be extracted from it.  Also, Bell inequalities
can be used as a simple test for the security of quantum
cryptography \cite{bellcrypto1}. The rough idea is that
past-generated correlations could have been created (and thus
known) by an eavesdropper, and so are not useful for cryptography,
whereas other kinds of  correlations cannot be created by the
eavesdropper and so are useful. The connection holds very closely
in some of the main cases of interest \cite{bellcrypto2}.

Bell inequalities are also related to classical communication
complexity \cite{commcomplexity}: how much communication do two
parties need in order to perform some joint task?  A final reason
is that no experiment has definitively demonstrated correlations
outside the past-generated set\cite{Aspectreview}.  We hope to one
day find an inequality which will allow us to do this without
waiting for improved technology.

A typical experiment to test for correlations begins by creating a
particular quantum mechanical state of two particles, and sending
one to site A, and the other to site B.  We then perform one of a
certain number, $m_A$ and $m_B$ say, of possible measurements,
$i_A$ and $i_B$, at each of the two sites, $A$ and $B$.  Each
measurement has a certain number, $n_A$ and $n_B$, of possible
outcomes, $j_A$ and $j_B$.  We then repeat the experiment many
times to get accurate probabilities for each set of joint
outcomes, $P(j_A,j_B | i_A,i_B)$.  The $d \equiv m_A m_B n_A n_B$
probabilities can be thought of as a point in a $d$ dimensional
space.  We are interested in the set of all points which can be
described using past-generated correlations. This set is convex,
and the boundary is defined by hyperplanes. It is straightforward
to list the $n_A^{m_A} n_B^{m_B}$ vertices. We would like to know
the faces, otherwise known as the Bell inequalities.  We want to
know how many different types of faces there are, and moreover,
which ones are relevant for quantum mechanics.

Characterising the set of past-generated correlations is
difficult.  Here difficult is meant in a technical sense. Suppose
we are given a point and asked if it is in the set. Finding the
answer is a finite computation, but will take more time with more
possible measurements, $m_A m_B$.  In fact, Pitowsky has shown
this problem to be NP-complete \cite{Pitowskybook}.  Furthermore,
suppose we are given an inequality, and wish to know whether or
not it is a face.  This problem is of similar difficulty: co-NP
complete \cite{Pitowskyfaces}.

Since the general problem is so hard, we calculated the Bell
inequalities for various small numbers of measurements and
outcomes. Surprisingly, we have found that for small numbers of
measurements and outcomes there are very few inequivalent Bell
inequalities. For the case $(m_A=2, m_B=2, n_A=2, n_B=2)$
(hereafter called $2222$), Fine \cite{Fine} has shown that up to
certain equivalences which we shall describe below, the only Bell
inequality is the CHSH \cite{CHSH}. For the case $3322$ we have
found that there is only a single new inequality. Considering the
complexity of the problem, this is very surprising and simple.
Furthermore, the inequality is relevant, since there are states
which violate it but do not violate the CHSH inequality. We
believe this is the first time that performing more than two
measurements has been shown to be useful for detecting non-local
correlations.

The confidence gained from such results (we shall describe more)
has also helped us to generalise this inequality to the case
$mmnn$.  Thus the computational approach, whilst limited, has
proved rather useful.  However we do not know if these
inequalities are relevant: are there states which do not violate
the previous inequalities which violate these new ones?
Computationally, it seems the answer is "no", though we have no
systematic method for checking this.  But if these inequalities
are not relevant, this would further reduce the number of
inequivalent, relevant Bell inequalities.

Before describing our results, we must describe more precisely the
set of correlations, and what we mean by equivalent and relevant
inequalities. As previously stated, the correlations live in a $d$
dimensional space with components representing $P(j_A,j_B |
i_A,i_B)$.  Some of these components are redundant, however. For
example, for any fixed measurement, there has to be at least one
outcome, ie. $\sum_{j_A,j_B} P(j_A,j_B | i_A,i_B)=1$. Also the
correlations we are interested in, those of quantum (and
classical) mechanics, do not allow faster than light signalling.
In other words, the distribution of outcomes on one particle does
not depend upon the choice of measurement made on the other:
$\sum_{j_A} P(j_A,j_B | i_A,i_B) = \sum_{j_A} P(j_A,j_B |
i'_A,i_B)$. Therefore we work in the subspace of dimension
\begin{equation}
d_2 \equiv m_A m_B (n_A-1) (n_B-1) + m_A (n_A-1) + m_B (n_B-1)
\end{equation}
which satisfies all of these constraints. Our subspace can be
labelled using the components $P(j_A,j_B | i_A,i_B)$ for
$j_A=0..(n_A-2)$, $j_B=0..(n_B-2)$, $i_A=1..m_A$, $i_B=1..m_B$;
the components $P(j_A | i_A)$ for $j_A=0..(n_A-2)$, $i_A=1..m_A$;
and the components $P(j_B | i_B)$ for $j_B=0..(n_B-2)$,
$i_B=1..m_B$.

We are interested in the faces of the convex set in this reduced
space. These are "tight" inequalities.  There are of course other
inequalities which are satisfied by all the points inside the
past-generated set, but such inequalities are less useful for
detecting non-local correlations.  An inequality will describe a
$d_2-1$ dimensional hyperplane, and so has $d_2$ components.

Some of these faces are equivalent.  For example, when defining
the experiment, we have to decide which outcome $j_A$ or $j_B$ is
which, and which measurement $i_A$ or $i_B$ is which, and which
particle is $A$ and which $B$.  Since these choices are arbitrary,
we shall consider two inequalities to be equivalent if they can be
converted into one another simply by relabelling these local
choices.

Having found the inequivalent faces, we would like to know if they
are violated by quantum mechanics.  As our motivation for studying
bell inequalities comes from quantum physics, we are interested
only in the faces which are violated.

Given two inequalities which are violated, we define the first to
be {\it non-redundant} if quantum mechanics gives a point
$P(j_A,j_B | i_A,i_B)$ which violates the first inequality , but
which does not violate the second inequality (or any inequality
equivalent to the second). Given several inequalities, we could
look for the minimal set of inequalities such that none are
redundant.

We are more interested in a classification which comes from
quantum mechanical states.  Given two inequalities we define the
first to be {\it relevant} if there exists a quantum state which
violates it for some choice of measurements, but does not violate
the second inequality for any choice of measurements. Similarly
for the second inequality. Given a set of inequalities, we want to
find the minimal set of relevant ones. Note that this will lead to
a minimal set no larger that the set of non-redundant
inequalities: in fact it may be much smaller.

One is often interested in quantum systems of a certain size, like
2-qubit systems.  How many relevant inequalities are there for
2-qubits?  Before the present work, only one - the CHSH - was
known to be relevant.  An open question was whether more
measurements (or more outcomes) could help. Here we show that
inequalities with 3 measurements are relevant. We do not know
whether 4 or more will help.

More outcomes may also be useful even on qubits, since we could
perform a several outcome POVM.  Whilst the usefulness remains an
open question, we can at least put an upper bound of $d^2$
outcomes for any measurement on a d-dimensional system.  Thus for
2-qubits, it is not useful to have more than 4 outcomes for any
one measurement.  The reason for this is that a POVM in a
$d^2$-dimensional space (the dimension of the density matrix) can
always be viewed as a classical probabilistic mixture of POVM's
with $d^2$ outcomes \cite{Parthasaraty}.  In other words, the many
outcome POVM can be viewed in two stages.  The first rolls an
independent dice to decide which few-outcome measurement to make.
The second performs the few outcome measurement, and gives the
appropriate outcome. Adding local randomness which is under the
control of the lhv model cannot add non-locality, and so any
non-locality present in such a many outcome measurement must
already be there in the few outcome measurements, and hence in a
few outcome inequality.

The limit on the number of useful outcomes is also interesting
since it suggests there are inequalities which are irrelevant for
2-qubits, but which are useful for higher dimensional systems. For
example the 5-dimensional CGLMP inequality \cite{CGLMP}, which
deals with the case $2255$, is irrelevant for qubits (it has too
many outcomes).  On the other hand, there are 5-dimensional states
which violate this inequality which are not known to violate any
lower dimensional inequalities.  With our present knowledge, this
inequality is indeed useful for such systems.

In order to find all the inequivalent inequalities, we have
several software tools.  The main one is a linear programming tool
which takes a list of vertices as input and, after some time,
outputs all the faces \cite{cdd}.  We have written a small matlab
program which, given $m_A, m_B, n_A, n_B$, produces a list of the
vertices.  The vertices are given by distributions which factor
into two local probability distributions, ie.
\begin{equation}
P(j_A,j_B | i_A,i_B)=P(j_A | i_A) P(j_B | i_B),
\end{equation}
and for which all the local probabilities are either 0 or 1, eg.
for $2222$,
\begin{eqnarray}
P(j_A=0 | i_A=1) & = & 0, \\
P(j_A=0 | i_A=2) & = & 1, \\
P(j_B=0 | i_B=1) & = & 1, \\
P(j_B=0 | i_B=2) & = & 0.
\end{eqnarray}
After we have the list of faces, we put this into a second matlab
program which removes the equivalences, leaving us with the
inequivalent inequalities. These software are all deterministic,
and so give the exact solution.  The bottleneck is the freely
available linear programming tool, which is optimized for certain
kinds of convex sets, but not for the equivalences which we have
here.  We have a final piece of software \cite{Bernardsoftware}
which, given an inequality and either the size of the quantum
system (eg. 2 qubits) or a specific quantum state,
probabilistically finds the maximum value of the inequality.

For the case $2222$ the software reproduces Fine's result that
there are only two types of inequality.  One is the trivial one
that probabilities are positive, ie. $P(j_A,j_B | i_A,i_B) \ge 0$.
This occurs $m_A m_B n_A n_B = 16$ times (to cover all the joint
probabilities).  There is no need for inequalities stating that
probabilities should be no greater than $1$, since this follows
from all the probabilities being positive.  The other type of
inequality is the CHSH, which occurs $8$ times. We write it here
in a form closer to that of the CH inequality \cite{CH})
\begin{eqnarray}
I_{CHSH} & = & P(A_1 B_1) + P(A_2 B_1) + P(A_1 B_2) - P(A_2 B_2) \nonumber \\
& & - (P(A_1)+P(B_1)),
\end{eqnarray}
where $P(A B)$ is the probability that when $A$ and $B$ are
measured we get the outcome $0$ for both measurements.  $I_{CHSH}
\le 0$ for lhv correlations.  Quantum mechanics can attain values
up to $\frac{1}{\sqrt{2}}-\frac{1}{2}$.

It will be useful for later on to write this inequality in the
following way:
\begin{equation}
I_{CHSH} = \left(
\begin{tabular}{c||cc}
  & -1 & 0 \\
 \hline \hline
-1 & 1 & 1  \\
0 & 1 & -1  \\
\end{tabular}
\right)
 ,
\end{equation}
where the table gives the coefficients we are to put in front of
the probabilities:
\begin{equation}
\left(
\begin{tabular}{c||cc}
 & $P(A_1)$ & $P(A_2)$ \\
 \hline \hline
$P(B_1)$ & $P(A_1 B_1)$ & $P(A_2 B_1)$  \\
$P(B_2)$ & $P(A_1 B_2)$ & $P(A_2 B_2)$  \\
\end{tabular} \right).
\end{equation}

Next we computed the case $2322$.  This is a choice between two
2-outcome measurements on one particle, and between three
2-outcome measurements on the other particle.  Here we found no
new inequalities.  We have only that the probabilities must be
positive, and CHSH inequalities where results of one the 3
measurements is ignored, eg.
\begin{eqnarray}
I'_{CHSH} & = & P(A_1 B_2) + P(A_2 B_2) + P(A_1 B_3) - P(A_2 B_3) \nonumber \\
& & - (P(A_1)+P(B_2)).
\end{eqnarray}
There are $m_A m_B n_A n_B=24$ "positive probability" faces, and
$8 \left(
\begin{tabular}{c}
3 \\
2 \\
\end{tabular} \right) =24$ CHSH faces - we have to choose 2 of the
three possible measurements for B, and once these are chosen we
have the 8 versions of the CHSH inequality which appear in the
$2222$ case.  This gives a total of $48$ faces.

 We have analytically extended this result to the case
$2m22$.  We find that there are no new inequalities for this case.
Our proof is essentially to note that the proof of Fine
\cite{Fine} for the $2222$ case extends naturally to the $2m22$
case.  Fine's proof works by starting with the measured
probabilities $P(j_A,j_B | i_A,i_B)$, which are assumed to satisfy
the CHSH inequalities.  He then constructs a lhv model which
reproduces the measured probabilities.  We shall describe the
construction for the $3222$ case: the general $m222$ case follows
very naturally.

First we define $\beta$ to be the minimum of 8 quantities:
$P(B_1)$, $P(A_1 B_1) + P(B_2) - P(A_1 B_2)$, and the other 6
quantities which come from exchanging $A_1$ for $A_2$ or $A_3$ and
$B_1$ for $B_2$ in the previous expressions.  We set $P(B_1,B_2)
\equiv \beta$.  $P(B_1)$ and $P(B_2)$ are experimentally
measurable so we can complete the distribution for $B_1$ and $B_2$
by $P(B_1, \bar{B}_2) \equiv P(B_1) - \beta$, $P(\bar{B}_1, B_2)
\equiv P(B_2) - \beta$, and $P(\bar{B}_1, \bar{B}_2) =
1-P(B_1)-P(B_2)+ \beta$, where $P(\bar{B}) \equiv P(B=1)$. One can
check that all these probabilities are positive, using the fact
that all the measured probabilities $P(j_A,j_B | i_A,i_B)$ are
positive.

We extend this to a lhv model for $A_1$, $B_1$ and $B_2$. We
define $\alpha$ to be the minimum of $P(A_1,B_1)$, $P(A_1,B_2)$,
$\beta$ and $\beta -
(P(A_1)+P(B_1)+P(B_2)-P(A_1,B_1)-P(A_1,B_2)-1)$. We set
$P(A_1,B_1,B_2) \equiv \alpha$.  We can check this is well defined
using the CHSH inequalities.  We complete the distribution for
$(A_1, B_1, B_2)$ using the quantities we already have.  ie.
\begin{eqnarray}
P(A_1,B_1,\bar{B}_2) & \equiv & P(A_1,B_1)-\alpha, \\
P(A_1,\bar{B}_1,B_2) & \equiv & P(A_1,B_2)-\alpha, \\
P(A_1,\bar{B}_1,\bar{B}_2) & \equiv &
P(A_1) - P(A_1,B_1) \nonumber \\ & & - P(A_1,B_2) + \alpha, \\
P(\bar{A}_1,B_1,B_2) & \equiv & \beta-\alpha, \\
P(\bar{A}_1,B_1,\bar{B}_2) & \equiv &
P(B_1)-P(A_1,B_1)-(\beta-\alpha), \\
P(\bar{A}_1,\bar{B}_1,B_2) & \equiv &
P(B_2)-P(A_1,B_2)-(\beta-\alpha), \\
P(\bar{A}_1,\bar{B}_1,\bar{B}_2) & \equiv &
P(A_1,B_1)+P(A_1,B_2)+(\beta-\alpha) \nonumber \\ & &
+1-P(A_1)-P(B_1)-P(B_2).
\end{eqnarray}
That these are all positive follows from the CHSH inequalities.

In a similar way, we make lhv models for the triple $(A_2, B_1,
B_2)$, and the triple $(A_3, B_1, B_2)$.

We finally extend this to a distribution for
$(A_1,A_2,A_3,B_1,B_2)$ by
\begin{eqnarray}
P(A_1,A_2,A_3,B_1,B_2) & \equiv & P(A_1 | B_1,B_2) P(A_2 |
B_1,B_2) \nonumber \\
& & *P(A_3 | B_1,B_2) P(B_1,B_2),
\end{eqnarray}
where $P(A_1 | B_1,B_2) = P(A_1,B_1,B_2) / P(B_1,B_2)$. This gives
us a well defined lhv distribution which reproduces all the
measured probabilities.

For three possible measurements on each side, the case $3322$,
Garg and Mermin \cite{GargMermin} have shown that the CHSH
inequalities are not the only faces of the classical polytope.
They found a point which satisfies all the CHSH inequalities, but
does not admit a lhv model.  A complete list of the faces have
been computed by Pitowsky and Svozil \cite{Pitowsky}. They found
684 faces. Removing equivalent inequalities, we find a single new
inequality. The 684 faces of the $3322$ polytope are made up as
$m_A m_B n_A n_B=36$ "positive probability" faces, $8 \left(
\begin{tabular}{c}
3 \\
2 \\
\end{tabular} \right)^2 =72$ CHSH faces, and 576 equivalent new faces.
The new face is
\begin{equation}
I_{3322} = \left(
\begin{tabular}{c||ccc}
  & -1 & 0 & 0 \\
\hline \hline
 -2 & 1 & 1 & 1 \\
 -1 & 1 & 1 & -1 \\
  0 & 1 & -1 & 0 \\
\end{tabular}
\right).
\end{equation}

This expression satisfies $I_{3322} \le 0$ for past-generated
correlations.  For quantum mechanics a numerical optimization
suggests that the maximum value is $0.25$.  This value can be
attained by the maximally entangled state
\begin{equation}
\label{maxent}
 \ket{\psi} = \frac{1}{\sqrt{2}}(\ket{0,1} -\ket{1,0}).
\end{equation}
The measurements all lie in a plane, so we denote their position
by a single angle, the angle which they make with the z-axis in
the Bloch sphere. $A_1 = 0$, $A_2=\frac{\pi}{3}$, $A_3=\frac{2
\pi}{3}$, $B_1=\frac{4 \pi}{3}$, $B_2=\pi$, and $B_3=\frac{2
\pi}{3}$.

The most interesting feature of this inequality is that there
exist states which violate it which do not violate the CHSH
inequality.  For example, consider the 2-qubit state
\begin{equation}
\sigma = 0.85 P_{\ket{\phi}} + 0.15 P_{\ket{0,1}},
\end{equation}
where $P_{\ket{\phi}}$ is the projector onto the state
$\ket{\phi}$, and
\begin{equation}
\ket{\phi} = \frac{1}{\sqrt{5}}( 2 \ket{0,0} + \ket{1,1}).
\end{equation}
One can check (using the Horodecki criterion \cite{Horodecki})
that this state does not violate the CHSH inequality.  However it
does violate the $3322$ inequality, giving a value $\sim 0.0129$.
The measurements for this violation are Von-Neumann measurements
in the directions $(\theta_{azim},\theta_{polar})$, where
$\theta_{azim}$ is the azimuthal angle with the z-axis, and
$\theta_{polar}$ is the polar angle in the x-y plane, when we set
the z-axis to be in the direction of $\ket{0}$), and the x-axis to
be in the direction of $\ket{0}+\ket{1}$:
\begin{eqnarray}
A_1 & = & (\eta, 0), \\
A_2 & = & (\pi - \eta, 0), \\
A_3 & = & (0, 0), \\
B_1 & = & (\pi - \chi, 0), \\
B_2 & = & (\chi, 0), \\
B_3 & = & (\pi, 0),
\end{eqnarray}
where $\cos \eta = \frac{1}{2 \sqrt{2}}$, and $\cos \chi =
\sqrt{\frac{7}{8}}$.

To compare the inequalities CHSH and $I_{3322}$, we numerically
calculated the maximum violation, $Tr(B \rho)$, of $I_{3322}$ for
all possible Von-Neumann measurements for a family of states
parameterised by $\theta$,
\begin{equation}
\label{3322states}
 \rho_{\theta} = \lambda_{CHSH} P_{cos \theta
\ket{0,0} + sin \theta \ket{1,1}} + (1-\lambda_{CHSH})
P_{\ket{0,1}},
\end{equation}
where $\lambda_{CHSH}$ is chosen so that each state in the family
$\rho_{\theta}$ gives the maximal value of the CHSH inequality
which can be obtained by lhv theories.  In order to give some
meaning to the size of the violation, we have re-scaled $I_{CHSH}$
and $I_{3322}$ so that the lhv maximum is $1$, and the maximally
mixed state $\rho=\frac{I}{4}$ gives the value $0$, ie.
$\tilde{I}_{CHSH}= 2 I_{CHSH}+1$, $\tilde{I}_{3322}=I_{3322}+1$.
The results are in FIG 1.
\begin{figure}[h]
\begin{center}
\epsfxsize=9cm \epsfbox{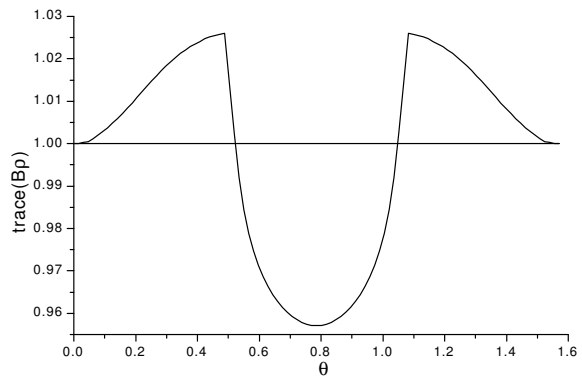}
 \caption{Maximum value of two Bell inequalities for a family of states.
 The straight horizontal line is that of $\tilde{I}_{CHSH}$, whilst the
 curve is for $\tilde{I}_{3322}$.}  \label{fig1}
\end{center}
\end{figure}

The software which calculates these maximum quantum mechanical
values converges quickly to very consistent results, giving us
confidence that they are correct.  We see that the new inequality
is most important not for states near the maximally entangled
state ($\theta=\frac{\pi}{4}$), but rather for states with less
symmetry.

It also seems that the CHSH inequality is still relevant: there
are states which violate it which do not violate $I_{3322}$.  This
is an illusion, due to the fact that we only maximized the
violation over non-degenerate Von-Neumann measurements.
Surprisingly, the maximum violation of the new inequality is often
given by degenerate measurements.  For example, if we take
$I_{3322}$ and set $A_3=1$ and $B_1=1$, the remaining measurements
give us the CHSH inequality.  Thus given $I_{3322}$, the CHSH
inequality is no longer relevant.

$I_{3322}$ shows a very direct non-locality in the states of
equation (\ref{3322states}). It is worth noting that such states
are also non-local by Popescu's "hidden non-locality" criterion
\cite{hidden}, \cite{hidden2}. In this one first makes local
filtrations to the state on both particles, and then performs a
standard CHSH test on the state which will emerge if both
particles pass the filters. One only looks at the data in the case
where both filters are passed, and if this data violates the CHSH
inequality we are assured that the original state is non-local.
For our states we would apply a local filter to particle $A$ which
lets state $\ket{1}$ pass, and absorbs state $\ket{0}$ with high
probability. We would simultaneously apply a local filter to
particle $B$ which lets $\ket{1}$ pass, and absorbs $\ket{0}$ with
high probability.  The idea is that each component of the
entangled state is only filtered once, whereas the noise term is
filtered twice.  If both particles pass the filter, the state is
very close to a pure entangled state.  Since all pure qubit states
violate the CHSH inequality \cite{pureviolate}, this one does too,
and we have shown hidden non-locality.

We have also tested the new inequality on the 2-qubit Werner state
\cite{Werner}:
\begin{equation}
\rho_p = p P_{\ket{\psi}} + (1-p) \frac{I}{4},
\end{equation}
where $\ket{\psi}$ is the maximally entangled state, as in eqn.
(\ref{maxent}). This state is interesting since despite being
entangled for $p>\frac{1}{3}$, Werner gave an explicit LHV model
for all Von-Neumann measurements for $p \le \frac{1}{2}$.
$I_{CHSH}$ is violated for $p > \frac{1}{\sqrt{2}}$, leaving a
region $\frac{1}{2}<p \le \frac{1}{\sqrt{2}}$ where there may or
not be model.  If there is not a model, it must be that some Bell
inequality $n_1 n_2 22$ is violated.  We find that $I_{3322}$
gives a violation only for $p > \frac{3}{4}$, suggesting that such
a model exists.

Another important feature of this inequality is that non-locality
can be shared between qubits.  Imagine that we have 3 qubits, $A$,
$B$ and $C$, and we ask whether one can simultaneously give
non-local correlations with the second (summing over the third
particle's outcomes), and non-local correlations with the third
(summing over the second particle's outcomes).  For CHSH
non-locality, the answer is "no" \cite{CHSHsharing}: non-locality
is monogamous. We can violate the inequality between parties $A$
and $B$, or $A$ and $C$, or have both pairs give the lhv maximum,
but never violate both at the same time.

The non-locality shown by the new inequality can be shared.  Take
the three qubit state
\begin{equation}
\ket{\psi}=\mu
\ket{000}_{ABC}+\sqrt{\frac{1-\mu^2}{2}}(\ket{110}_{ABC}+\ket{101}_{ABC}),
\end{equation}
with $\mu=0.852$.  Qubits $B$ and $C$ are symmetric, and qubits
$A$ and $B$ violate $I_{3322}$ giving a value $0.0041$. The
measurements are defined by the azimuthal and polar angles:
\begin{eqnarray}
A_1 & = & (\alpha, 2 \pi-\beta), \\
A_2 & = & (\alpha, \pi-\beta), \\
A_3 & = & (\frac{\pi}{2}, 2 \pi-\delta), \\
B_1 & = & (\gamma, \pi+\delta), \\
B_2 & = & (\gamma, \delta), \\
B_3 & = & (\frac{\pi}{2}, \beta),
\end{eqnarray}
where $\alpha=2.8252$, $\beta=0.1931$, $\delta=0.0804$ and
$\gamma=2.5445$.

Inspired by this case, we have found a generalization of this
inequality to the $mm22$ case.  For $4422$ it looks as follows:
\begin{equation}
I_{4422} = \left(
\begin{tabular}{c||cccc}
   & -1 & 0 & 0 & 0 \\
  \hline \hline
  -3 & 1 & 1 & 1 & 1 \\
  -2 & 1 & 1 & 1 & -1 \\
  -1 & 1 & 1 & -1 & 0 \\
  0 & 1 & -1 & 0 & 0 \\
\end{tabular}
\right).
\end{equation}
The past-determined correlations are always $\le 0$. The
generalisation to $mm22$ should now be clear.  The main part of
the matrix has entries $1$ in every position from the top left
corner to the backwards diagonal.  There is then one backwards
off-diagonal line of $-1$ entries, and afterwards $0$'s complete
the matrix.  We then subtract $\sum_{i=1}^{m_B} (m_B-i)
P(B_i)+P(A_1)$.

We shall prove the lhv maximum by induction.  Starting from the
lower left corner, we can see that $I_{mm22}$ contains all the
inequalities from the same family with less measurements. If we
just take measurements $A_1$ and $B_m$ (ignoring the other
measurements by setting their outcomes to $1$), we have a positive
probability face. Adding measurements $A_2$ and $B_{m-1}$ gives
the CHSH inequality. Adding $A_3$ and $B_{m-2}$ gives us
$I_{3322}$.  Let us assume that we have proved $I_{(m-1) (m-1) 2
2} \le 0$.  To get a value larger than $0$ for $I_{mm22}$, we must
total at least $+1$ in the terms which were not present in the
previous inequality.  We can only do this by setting $A_i=0
\forall i$, and $B_1=0$.  This gives us $+1$ in the new terms. But
now we have a $-1$ from $P(A_1)$. Whatever we put for the values
of $B_j$ for $j=2..m$, each row $j$ contributes exactly $0$ to the
total, giving us $I_{mm22}=0$, and proving that this is the
maximum.

Unfortunately, we do not know if $I_{mm22}$ is a face for all $m$.
We have found computationally that it is indeed a face for $m \le
7$, and suspect that this will generalise.  We also do not know if
any of the other inequalities $mm22$ are relevant after one
already has the $3322$ inequality.  An analytic problem here is
that we have no simple criterion to say which states definitely do
not violate the $3322$ inequality.  Even computationally we have
not yet found an example of a state which would violate one of the
inequalities $mm22$ without violating $I_{3322}$.  Whilst one
would expect to find new, inequivalent faces at every $m$, it is
not clear that they will all be relevant, particularly if we fix
ourselves to a certain quantum system size, like 2-qubits.

For the case $3422$, we find 12480 faces, which include three new
inequalities, along with $I_{3322}$, the CHSH and positive
probability faces. As for our $I_{mm22}$ inequalities, we do not
know a good way to discover whether these new inequalities are
relevant, or to uncover other interesting features they may
possess.  The new inequalities are in Appendix A.

We have not gone beyond $3422$ in a complete way at present since
our software takes too long to run on our PC. We are able to
produce a subset of the faces, but have not investigated this
direction.

We can also look at inequalities with more measurement outcomes.
For $2223$ there are no new types of face.  There are only the
CHSH, and positive probability faces.  To use the CHSH inequality
(which is defined for two outcome measurements) for three outcome
measurements, we map the three outcomes into two effective
outcomes by putting two of the original outcomes together.  We
then put the effective outcomes into the CHSH inequality. This can
be done in $3$ different ways for each of particle B's
measurements.  Since there are $8$ versions of the CHSH inequality
for $2222$, this gives $8*3^{m_B}=72$ faces.  There are $m_A m_B
n_A n_B=24$ positive probability faces, making $96$ faces in
total.

For $2224$ we again find no new types of faces.  There are $32$
positive probability faces, and $392$ CHSH faces.  Note that there
are two different ways to put together the $4$ outcomes: we can
group $3$ of them together against the 4th, or put them in $2$
groups of $2$.  There are $4$ ways to do the first, and $3$ to do
the second, giving $8*(4+3)^{m_B}=392$ faces.

For $2225$ and $2226$ we have computed a list of all the faces,
but have not been able to sort them.  The number of faces, $1840$
and $7736$, is that which one predicts assuming there are no new
faces.  Therefore we conjecture that there are no faces beyond the
CHSH for the case $222n$.

For the case $2233$, there is only one new type of inequality,
which is already found in \cite{3CH,CGLMP}. This can be written as
\begin{equation}
\label{I2233}
 I_{2233} = \left(
\begin{tabular}{c||cc|cc}
  & -1 & -1 & 0 & 0 \\
  \hline \hline
  -1 & 1 & 1 &  0 & 1 \\
  -1 & 1 & 0 &  1 & 1 \\
  \hline
  0 & 0 & 1 &  0 & -1 \\
  0 & 1 & 1 & -1 & -1 \\
\end{tabular}
\right).
\end{equation}

The columns of the correlation part of the matrix correspond to
$A_1=0$, $A_1=1$, $A_2=0$ and $A_2=1$.  The rows are in the same
order, for particle B.  Thus the first entry is $P(A_1=0,B_1=0)$.
For lhv models $I_{2233} \le 0$.

The total number of faces for the case $3322$ is $1116$, of which
$36$ are positive probability, $8*3^{m_A}3^{m_B}=648$ are CHSH,
and $432$ are $I_{2233}$.

For states of 2 qutrits of the form
\begin{equation}
\rho_p = p P_{\ket{\psi}} + (1-p) \frac{I}{9},
\end{equation}
where
$\ket{\psi}=\frac{1}{\sqrt{3}}(\ket{0,0}+\ket{1,1}+\ket{2,2})$,
$I_{2233}$ is violated by states with more noise (a smaller $p$)
than the CHSH inequality.  Thus it is relevant.  On the other
hand, we can recover the CHSH inequality from this one by using
the outcomes $1$ and $2$ for measurements $A_1$ and $A_2$, and
outcomes $0$ and $2$ for measurements $B_1$ and $B_2$.  Once we
have this new inequality, the CHSH is no longer relevant.

This inequality has been generalised to $22nn$ \cite{CGLMP}. The
generalised inequalities are known to be faces for all $n$, and to
be the only faces which exist of a certain form \cite{Lluis}.  In
our present notation, they look simpler than they did in the
original paper, for example
\begin{equation}
\label{I2244}
 I_{2244} = \left(
\begin{tabular}{c||ccc|ccc}
  & -1 & -1 & -1 & 0 & 0 & 0\\
  \hline \hline
  -1 & 1 & 1 & 1 & 0 & 0 & 1 \\
  -1 & 1 & 1 & 0 & 0 & 1 & 1 \\
  -1 & 1 & 0 & 0 & 1 & 1 & 1 \\
  \hline
  0 & 0 & 0 & 1 & 0 & 0 & -1 \\
  0 & 0 & 1 & 1 & 0 & -1 & -1 \\
  0 & 1 & 1 & 1 & -1 & -1 & -1 \\
\end{tabular}
\right).
\end{equation}
The lhv maximum is 0.  To see this, note that to get more, the
local terms $-P(B_1 \neq n)$ and $-P(B_1 \neq n)$ force us to try
to get $+1$ from the three pairs of measurements $(A_1,B_1)$,
$(A_1,B_2)$, and $(A_2,B_1)$.  But if we do this, we are forced to
get a $-1$ from $(A_2,B_2)$, leaving us with a total of $0$.

The generalisation of the inequality to more outcomes is as one
would guess.

For $2234$ we have computed all the faces, but not sorted them.
The total number of faces, $19128$, matches the number one expects
assuming there are no new inequalities.  Beyond this our program
would take too long to compute the complete solution.

What about inequalities combining more measurements and more
outcomes?  Garg and Mermin \cite{GargMermin2} have evidence which
suggests that such inequalities exist.  They found a quantum state
and measurements for the case $3333$ for which the results satisfy
all $2222$ and $2233$ inequalities, but are nevertheless
non-local. We have found a family of inequalities for the case
$mmnn$, which is a generalisation of the inequalities for $22nn$
and $mm22$. The first member is
\begin{equation}
\label{I3333}
 I_{3333} = \left(
\begin{tabular}{c||ccc}
 & -\bf{1} & \bf{0} & \bf{0} \\
\hline \hline
 -\bf{2} & X &  X & Y \\
 -\bf{1} & X &  Y & -Y \\
  \bf{0} & Y & -Y & -Z \\
\end{tabular}
\right),
\end{equation}
where $X = \left( \begin{tabular}{cc}
   1 & 1 \\
   1 &  0  \\
\end{tabular}
\right)$, $Y = \left( \begin{tabular}{cc}
   0 & 1 \\
   1 &  1  \\
\end{tabular}
\right)$, $Z = \left( \begin{tabular}{cc}
   0 & 1 \\
   0 &  0  \\
\end{tabular}
\right)$, and $\bf{C}$ is a row (or column) in which every entry
is C.

 $I_{3333} \le 0$ for past-generated correlations.  $X$
and $Y$ are the same matrices which appear in $I_{2233}$, and the
arrangement of $X$ and $Y$ is similar to the arrangement of the
elements of the matrix in $I_{3322}$.  $Z$ is a new matrix we put
in by hand, because the more natural matrix full of $0$'s did not
give us a face.  The local probabilities which we subtract are a
natural generalisation of the terms from $I_{2233}$ and
$I_{3322}$.

To generate the complete $mmnn$ family we first generalise the
number of measurements, then the number of outcomes. Adding one
more measurement gives
\begin{equation}
\label{I4433}
 I_{4433} = \left(
\begin{tabular}{c||cccc}
  & -\bf{1} & \bf{0} & \bf{0} & \bf{0} \\
  \hline \hline
  -\bf{3} & X &  X & X & Y \\
  -\bf{2} & X &  X & Y & -Y \\
  -\bf{1} & X & Y & -Y & -Z \\
   \bf{0} & Y & -Y & -Z & -Z \\
\end{tabular}
\right).
\end{equation}
The generalisation to $mm33$ follows a similar pattern to that for
$mm22$.  The matrix consists of entries $X$ for all the elements
from the top-left corner to just before the main backwards
diagonal. The main backwards diagonal has entries $Y$, and the
next backwards diagonal has entries $-Y$. The lower right corner
is filled by entries $-Z$. We then subtract $P(A_1 \neq
2)+\sum_{j=1}^{m_B} (m_B-j) P(B_j \neq 2)$.

To generalise to more outcomes, one only has to change the
matrices $X$, $Y$, and $Z$.  $X$ and $Y$ change exactly as they do
in the family $22nn$.  $Z$ grows in a slightly odd looking manner:
\begin{equation}
Z = \left( \begin{tabular}{cccc}
   0 & 0 & 0 & 1 \\
   0 & 0 & 1 & 1 \\
   0 & 1 & 1 & 1 \\
   0 & 0 & 0 & 0 \\
\end{tabular}
\right),
\end{equation}
which is the same as $Y$ but for the last row.

$I_{mmnn} \le 0$ for lhv theories. To prove this, we combine the
proofs of $I_{mm22} \le 0$ and $I_{22nn} \le 0$.  Starting at the
bottom left corner, and successively adding pairs of measurements,
we see that $I_{mmnn}$ contains all the inequalities $I_{m'm'nn}$,
with $m' \le m$.  For $m=1$, the inequality is trivial.  For
$m=2$, the inequality is $I_{22nn}$, which we have already proved.
For $m=3$, to get more than $0$ we need to get a positive
contribution from the terms added on after $m=2$.  Thus we must
get a $+1$ from all the combinations $(A_1,B_1)$, $(A_2,B_1)$,
$(A_3,B_1)$. But now we have $-1$ from $A_1$.  To get a total of
more than $0$, we need to find a row $B_k$ where the contribution
from that row is $+1$.  To see that this is impossible, first look
at the row $B_2$.  We want to pick up $+1$ from $(A_1,B_2)$ and
$(A_2,B_2)$ without picking up a $-1$ from $(A_3,B_2)$.  But the
$+1$'s in $(A_2,B_2)$, $(A_2,B_1)$ and $(A_3,B_1)$ force the $-1$
to occur, making the maximum of the row 0.  Otherwise look at row
$B_3$. Here setting $B_3=n-2$ is no use, since we get a $+1$ and a
$-1$. $B_3=n-1$ is clearly useless, also giving us $0$.  For $B_3
< n-2$, the matrix $-Z$ looks like $-Y$, and so will always give
$-1$ for $(A_3,B_3)$ when we get $+1$ for $(A_1,B_3)$, $(A_1,B_1)$
and $(A_3,B_1)$.  So we have proven $I_{33nn} \le 0$.  For more
measurements, a similar argument leads to a proof by induction.

We know computationally that these inequalities are faces for
$m=2$, $n \le 7$, for $m=3$, $n \le 6$, for $m=4$, $n \le 4$, and
for $m=5$, $n=3$.  We suspect this is true for all $m$ and $n$. As
was the case for our family of $mm22$ inequalities, we do not know
if these new $mmnn$ inequalities are relevant.

In summary we have found that for small numbers of measurements
and outcomes there are very few inequivalent Bell inequalities.
For the case $2m22$ there is only the CHSH inequality. We believe
the same to be true for the case $222n$.  For the cases $2233$ and
$3322$ there is the CHSH inequality, but only one other inequality
in each case. The new inequalities are relevant: they are violated
by states which do not violate the CHSH inequality.  The $3322$
case shows that the CHSH inequality is not the only useful one for
$2$-qubits: more measurements really help!  We have also
discovered a family of inequalities for the case $mmnn$, but do
not know if any of these inequalities are relevant.

Having found that there are remarkably few relevant inequivalent
faces, we have more time to study closely the ones which do exist.
Will some of them help us to perform an experiment definitively
ruling out lhv correlations in the lab?  Is there a close
connection between the new inequalities and a particular quantum
cryptography protocol? How does this compare with different types
of quantum correlations?  All these possibilities would be
interesting, but here we have a more fundamental message. The set
of past-generated correlations is not as complicated as we
thought.

\vspace{0.1cm}

{\bf Thanks to:} A. Acin, N. Brunner, R. Gill, N. D. Mermin, I.
Pitowsky, V. Scarani and A. Shimony for helpful comments and
suggestions, and S. Fasel for software support.

We acknowledge funding by the Swiss NCCR, "Quantum Photonics" and
the European IST project RESQ.

{\bf Note Added:} C. \'{S}liwa has independently found some of the
results contained in this paper.  Firstly that for the case $2n22$
there are no inequalities beyond the CHSH.  Secondly that for the
case $3322$ there is only a single new inequality (this is, like
our result, an exact computational result).  He has also
investigated the three party case $222222$.

\appendix
\section{New 3422 Inequalities}

There are 12480 faces for the case 3422, which include 3 new
faces.  There are $3*4*2*2=48$ positive probability faces,
$8
\left(
\begin{tabular}{c}
3 \\
2 \\
\end{tabular} \right)
\left(
\begin{tabular}{c}
4 \\
2 \\
\end{tabular} \right)=144$ CHSH faces, and $576
\left(
\begin{tabular}{c}
4 \\
3 \\
\end{tabular} \right)=2304$ $I_{3322}$ faces.  Then the three new
inequalities.  There are 2304 versions of
\begin{equation}
I_{3422}^1 = \left(
\begin{tabular}{c||ccc}
  & 1 & 1 & -2 \\
\hline \hline
 1 & -1 & -1 & 1 \\
 0 & -1 & 1 & 1 \\
  0 & 1 & -1 & 1 \\
 1 & -1 & -1 & -1 \\
\end{tabular}
\right),
\end{equation}
which has a lhv maximum of $2$. There are 3027 versions of
\begin{equation}
I_{3422}^2 = \left(
\begin{tabular}{c||ccc}
  & 0 & 1 & -1 \\
\hline \hline
 -1 & -1 & 1 & 1 \\
 0 & 0 & -1 & 1 \\
  -1 & 1 & 0 & 1 \\
 1 & -1 & -1 & 0 \\
\end{tabular}
\right),
\end{equation}
with a lhv maximum of 1. Finally 4608 versions of
\begin{equation}
I_{3422}^3 = \left(
\begin{tabular}{c||ccc}
  & 1 & 0 & -1 \\
\hline \hline
 0 & -2 & 1 & 1 \\
 0 & 0 & -1 & 1 \\
  -1 & 1 & 1 & 1 \\
 2 & -1 & -1 & -1 \\
\end{tabular}
\right),
\end{equation}

with a lhv maximum of 2.

\end{document}